\documentstyle[aps,floats,epsf]{revtex}
\begin{document}
\setlength{\baselineskip}{4.2mm}
\draft
\preprint{STH-2001-01}
\title{
\vspace{37mm}
Periodicity Manifestations in the Non-Locally Coupled Maps\\} 
\author{Tokuzo Shimada and Shou Tsukada\\}
\address{Department of Physics, School of Science and Technology, Meiji University\\
Higashi-Mita 1-1-1, Tama, Kawasaki, Kanagawa 214-8571, Japan\\} 
\maketitle
\begin{abstract}\\ \indent
We study how periodicity manifestations recently found in the turbulent globally 
coupled maps depend on the global feature of the couplings. 
We examine three non-locally coupled map models. 
In the first two, the all-to-all interaction is maintained but the coupling decreases 
with distance in a power and an exponential law. 
In the third, the interaction is uniform but cut off sharply. 
We find that, in all three and in dimension $D=1,2,3$, 
periodicity manifests universally from turbulence 
when the same suppression of the local mean field fluctuation 
is achieved by the non-local averaging.\\
\end{abstract}

\noindent
{\bf 1. Introduction}\\ 

The globally coupled map lattice (GCML) is one of the basic models of complex systems
with interacting chaotic elements.
The simplest GCML consists of $N$ identical logistic maps 
with uniform couplings via their mean field $h$.   
It has only two parameters, the non-linearity $a$ of element map 
and the coupling $\varepsilon$ of the averaging interaction. 
Yet it exhibits a rich variety of phases under the balance between the randomness 
and the coherence \cite{ka} and provides us with a simple testing ground  
of basic features of wide variety of coupled random elements
such as the Josephson junction array, vortex dynamics, multimode lasers as well as biological networks.
In particular, in its `turbulent regime'  --- the region of high non-linearity and very weak coupling--- 
the system at large $N$ exhibits intriguing properties in evolution. 

Firstly, the maps are under unfailing weak coherence even if  the parameters are 
so chosen that no visible synchronization of maps occurs \cite{kb1}.
As a result the mean square deviation (MSD) of the time series of mean field $h_t$  
is sizably enhanced from the value predicted 
by the law of large numbers  ($\overline{\delta h^2} \propto1/N$), 
while the $h_t$ distribution is Gaussian following the central limit theorem \cite{kb1}.
This was called as a hidden coherence and triggered much scrutiny \cite{sinha,pikovskyb,kc}.
Furthermore, it has been recently found that the periodic windows of the element map systematically 
foliate and produce various periodicity manifestations (PM's)
\cite{ts,pre,parravano,shibata}. 
Therefore, GCML in the turbulent regime is a system which sensitively mirrors the periodic windows 
with the background hidden coherence. 
The amazing fact that even at very weak coupling the system easily form periodic cluster attractors 
may have important consequences in a complex system of coupled chaotic elements, 
especially in the activity of the brain. 

In this Letter we study to what extent the PM's depend 
on the all to all uniform coupling feature of GCML. We will show that 
they occur universally in three non-locally coupled map models and  
their strength can be predicted by a simple estimation of the fluctuation 
of local mean fields around the overall mean field.

The PM's are organized by the maps by synchronization when the parameters 
$a,\varepsilon$ are in a balance that allows a reduction of the high $N$-dimensional 
dynamics to that of the element logistic map in a periodic window at non-linearity $b$.  
The balance defines foliation curves on the $(a,\varepsilon)$ plane 
and all GCML on the curves are universally governed by the window dynamics. 
The most prominent PM's are induced by the period three window \cite{ts}.  
If GCML is on the foliation curves of the $p3$ window, the maps organize themselves into 
almost equally populated three clusters, which oscillate mutually in period three --- $p3c3$ 
maximally symmetric cluster attractor (MSCA) \cite{pre}. 
With slightly higher $\varepsilon$ at the same $a$, that is, on the foliation curves 
from the intermittent region, the maps organize themselves in $p3c2$ cluster attractor. 
The $p3$ attractors are formed at any reduction factor $r$ ($r\equiv b/a \le 1)$.
Generally, for a small reduction ($r \gtrsim 0.95$), maps form $p=c$ MSCA and $p>c$ clusters respectively 
along the curves of period $p$ window and in the nearby higher $\varepsilon$.
In MSCA, the MSD of the mean field fluctuation is minimized due to the high population symmetry 
and all observed MSCA's are linearly stable \cite{pre}.      
Contrarily, in $p>c$ states, the MSD turns out extremely high due to missing clusters to fulfill the orbits.
For a large reduction ($r \lesssim 0.95$), the clusters are no longer formed
but their remnants induce the same structure in the MSD ---
a valley and peak respectively along the curves of a window and in the nearby higher $\varepsilon$.
We use below the MSD curve as a representation of PM's.\\

%
\noindent
{\bf 2. Non-locally coupled map models}

The GCML on the lattice $\Lambda$ is 
\begin{eqnarray}
   x_{P}(t+1)=(1-\varepsilon) f(x_{P}(t)) + \varepsilon h_t, ~P\in \Lambda 
\label{gcmlevolution}
\end{eqnarray}
with the mean field $h_t \equiv (1/N) \sum_{Q \in \Lambda} f(x_{Q}(t))$
and $f(x)=1-a x^{2}$.
This is a two step process; the independent mapping followed by an interaction 
via the mean field $h_t$ with an overall coupling $\varepsilon$.
By adding (\ref{gcmlevolution}) over $P$
we find a relation $
	\frac{1}{N}\sum_{P \in \Lambda} x_{P}^\prime 
          =  \frac{1}{N}\sum_{P \in \Lambda} f(x_{P})$---
{\it the mean field is kept invariant in the interaction}.
All the non-local models we consider below respect this invariance rule.\\ \\

\noindent
{\it 2.1. A power law model: POW$_\alpha$}\\

As an extension of GCML let us consider a model  
\begin{eqnarray}
x_{P}^\prime &=& (1-\varepsilon) f(x_{P}) +\varepsilon h_P,  \nonumber \\
h_P &\equiv& \sum_{Q \in \Lambda} W_{PQ} f(x_{Q}) \nonumber\\
&=&  c^{(\alpha)} f(x_{P}) +d^{( \alpha )} \sum_{\rho=1}^{\rho_{\max}} \frac{1}{\rho^{\alpha}} 
 \sum_{Q \in \Lambda_{\rho}(P)} f(x_{Q}), 
\label{genericevolution}
\end{eqnarray}
where each map couples to other maps via a {\it local mean field} $h_P$.
The $\Lambda_\rho (P)$ is a set of maps at an equal distance $\rho$ from a site $P$.
For simple analytic estimates below, we approximate it 
by a set of points on the boundary of a $(2 \rho +1)^D$ square (cube) for $D=2(3)$.
The number of maps in $\Lambda_\rho$, $n_\rho$, is then $2,~8 \rho,~24 \rho^2 +2$ 
 respectively for $D=1,2,3$. 
We impose the periodic boundary condition and the maximum `radius' $\rho_{\max}$ of 
$\Lambda_\rho$ is $(N^{1/D}-1)/2$.
As a requisite the weights for a local mean field $h_P$ must add to one. This constrains  
the coefficients $c^{(\alpha)}$ and $d^{(\alpha)}$ as
\begin{eqnarray}
c^{(\alpha)} + d^{(\alpha)} S^{(\alpha)}=1, ~\text{with}~ 
S^{(\alpha)}\equiv \sum_{\rho=1}^{\rho_{\max}} n_\rho / \rho^\alpha. \label{weightsumrule} 
\end{eqnarray} 
Under this constraint, one obtains with a simple arithmetic
both the above invariance rule and a further important relation
$\frac{1}{N}\sum_{P \in \Lambda} h_{P} = h $---
{\it the average of the local mean fields is nothing but the mean field of the whole system}.

\begin{figure}[t]
\begin{center}
\leavevmode
\epsfxsize=86mm
\epsfysize=74mm
\epsfbox{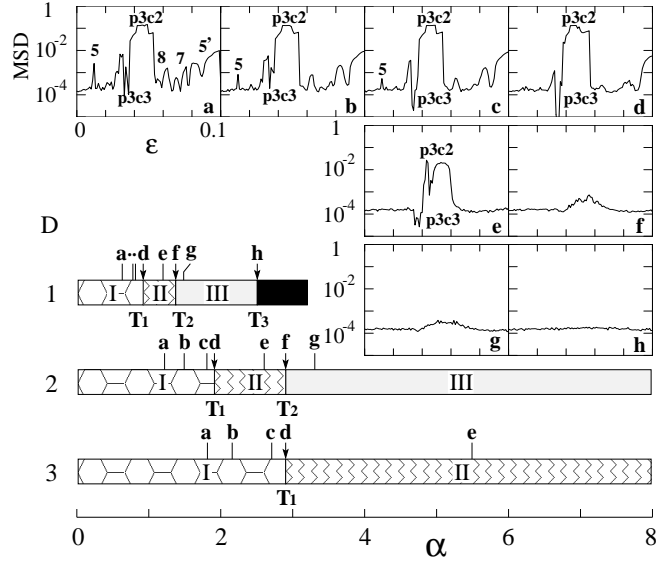}
\vspace{5mm}
{\footnotesize\caption{
The bar chart for the PM's in POW$_\alpha$ with $a=1.90$
and the MSD curves at the eight marking points (a-h).
I: full PM's, II: only the $p=3$ PM's,  III: only the hidden coherence. 
\label{barchartpow}}}
\end{center}
\end{figure}

Let us make (\ref{genericevolution}) into a model which interpolates the GCML and the 
nearest neighbor CML.
In order to match with GCML at $\alpha=0$, the coefficient must be
$ c^{(0)}=d^{(0)}=1/N$.
In order to match with the nearest neighbor CML
\begin{eqnarray}
 x^\prime_P = f(x_P) + \frac{\varepsilon}{n_1+1} 
            \left( \sum_{Q \in \Lambda_1 (P) }f(x_{Q}) - n_1  f(x_P)   \right)
\label{cml1}
\end{eqnarray}
 at $\alpha \rightarrow \infty$,  the coefficients must be $c^{(\infty)}=d^{(\infty)}={1}/({n_1+1})$.
In both limits, $c=d$. Therefore, 
we set $c^{(\alpha)}=d^{(\alpha)}$ for all $\alpha$ as the simplest interpolation. 
Normalizing the couplings by (\ref{weightsumrule}), 
we obtain a one parameter extension of GCML, POW$_\alpha$, with $h_P$ given by
\begin{eqnarray}
        \frac{1}{1+ S^{(\alpha)}} 
\left(
 f(x_{P}) 
	+ \sum_{\rho=1}^{\rho_{\max}} \frac{1}{\rho^\alpha} \sum_{Q \in \Lambda_{\rho}(P)} f(x_{Q}) 
\right).
\label{POW}
\end{eqnarray}
As shown in Fig.~\ref{barchartpow} PM's in POW$_\alpha$ diminish 
with the increase of $\alpha$. 
We choose eight marking points (a-h) and in the insets we display corresponding 
MSD curves. By the strength of PM's the $\alpha$ interval may be divided into three regions.
(I) --- One can observe all PM's that occur in GCML. 
From the GCML limit ($\alpha=0$) up to the point a, the full strength PM's are produced.  
The relevant dominant windows are marked on the MSD curve at a,
which agrees precisely with the curve in GCML \cite{pre}. 
From a to d, the peak valley structures, except for that due to $p3$
clusters, gradually diminish. The peak due to $p5$ window starts
diminishing at a and it becomes half-height at b. At c all the 
sub-dominant peak-valley structures vanish, and even the $p5$
peak vanishes at d. 
(II) --- The region of $p3$ PM's only. It starts from d
and the $p3c2$ peak disappears at e.  
At f, only a broad MSD peak is seen in the MSD curve. 
(III) --- Essentially the region of the hidden coherence.
Only broad MSD peak can be seen around the foliation zone of the 
$p3$ window.  At the start of III and near the top of the peak,
the correlator of maps decreases exponentially in time in a $p3$ motion.
At g, the broad MSD enhancement becomes half-height and 
the correlator fails to sense the periodicity everywhere. 
At h, the MSD enhancement disappears.
The transition points T$_1$, T$_2$, T$_3$ between 
the regions are d, f, h respectively. 
As we see in the bar chart, the variation of PM's occurs most quickly
in $D=1$ and it is prolonged in higher dimensions.
We note that $\alpha_{T_1} \approx 0.9,~1.9,~2.9$, 
approximately in the ratios $1:2:3$ for $D=1, 2, 3$ respectively.\\ \\

\noindent
{\it 2.2. A coupled map lattice with exponentially decaying couplings: EXP$_{\rho_0}$\\}

Similarly,  we obtain a model with $h_P$ given by
\begin{eqnarray}
  \frac{1}{1+ S^{(\rho_0)}} 
\left( f(x_{P}) 
+ \sum_{\rho=1}^{\rho_{\max}}w_{\rho, \rho_0} \sum_{Q \in \Lambda_{\rho}} f(x_{Q}) \right),  \label{EXP}
\end{eqnarray}
where $w_{\rho, \rho_0}=\exp(-({\rho -1})/{\rho_0})$ is the exponentially decaying coupling 
and $S^{(\rho_0)}= \sum_{\rho=1}^{\rho_{\max}} n_\rho w_{\rho, \rho_0}$.
This reduces to GCML at $\rho_0\rightarrow \infty$ and the nearest neighbor CML at $\rho_0 \rightarrow 0$.
The PM's diminish with {\it decreasing} $\rho_0$ via the same process as above.\\ \\

\noindent
{\it 2.3. A coupled map lattice with an interaction range $\kappa$: CML$_\kappa$}\\

\begin{figure}[t]
\begin{center}
\leavevmode
\epsfxsize=91.7mm
\epsfysize=65.5mm
\epsfbox{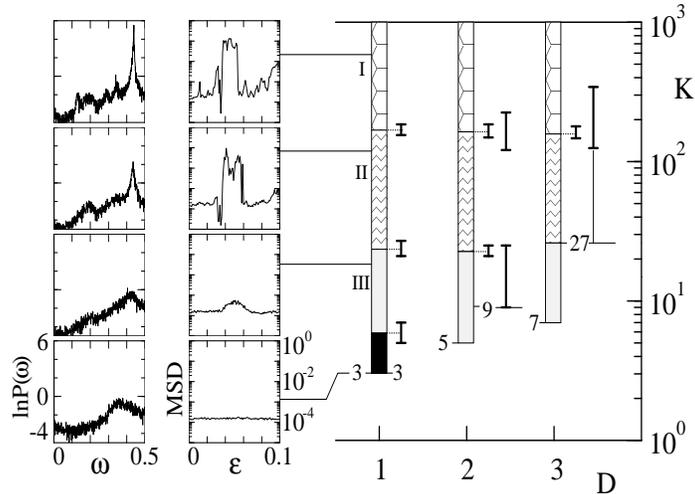}
\vspace{5mm}
{\footnotesize\caption{
Right: Bar chart of PM's in CML$_\kappa$ at $a=1.90$.  
I: full PM's, II: only the $p=3$ PM's, III: only the hidden coherence. 
The error bars compare the ambiguities in determining the transition points 
by two types of neighborhood.  
Center: The MSD curves at the pointed positions.
Left: Power spectrum of the mean field at $(a, \varepsilon)=(1.99,0.10)$ in $D=1$.
\label{barchartcmlk}}}
\end{center}
\end{figure}

Above two models maintain all-to-all coupling feature of GCML. 
Let us now consider a non-local CML with $h_P$ given by
\begin{eqnarray}
\frac{1}{K} \left( f(x_P) + 
\sum_{\rho=1}^\kappa
\sum_{Q \in \Lambda_\rho (P) } f(x_{Q}) \right).  \label{cmlkappa}
\end{eqnarray}
Here $K=(2 \kappa+1)^D$ is the number of maps within a range $\kappa$.
We find that the PM's diminish with {\it decreasing} $\kappa$, again via the same patterns of MSD curves. 
Furthermore, we find that remarkably the same PM's occur irrespective to the dimensions if the 
neighborhood encloses the same number of maps. 
For instance,  the range $\kappa$ at T$_1$ is $77-92, 5-6, 2-3$
in $D=1,2,3$ respectively,  but  the number of maps $K$ 
within  $\kappa$ is $155-185$, $121-169$, $125-343$ in $D=1,2,3$.  
The large error in $D=3$ comes from the large-step increase of $K$ with $\kappa$. 
Thus we determine the marking points in CML$_\kappa$  
by a refined neighborhood; a set of lattice points $Q$ around $P$ with 
$\sum_{i=1}^D (\Delta \rho_i)^2 \le \kappa^2$.
In Fig.~\ref{barchartcmlk} we show the three regions by a bar chart in terms of $K$. 
We find the bars in $D=1,2,3$ agree each other well.
CML$_\kappa$ was used in an analysis of hidden coherence from the view of `beat' of 
mean field \cite{sinha}. The Fourier power spectrum of $h_t$ 
is shown also in Fig.~\ref{barchartcmlk}. Interestingly,  the Fourier peaks due to the beat become 
outstanding in accord with the onset of PM's.\\ \\

\noindent
{\bf 3. The universality of PM's in non-local models}\\

The difference between GCML and other non-local models is only in the interaction step.
In GCML, the maps contract uniformly to $h$ by a factor $1-\varepsilon$, 
while in others a map $f(x_P)$ is contracted to the local mean field $h_{P}$
which distributes around the overall system mean field $h$. 
Therefore, when the variance of $\xi_P\equiv h_{P} - h$ is large, some distortion of map configuration must unavoidably be introduced in the interaction step. 
Contrarily, when the variance is small, such a distortion will be avoided and the non-local 
system may evolve just in the same way with GCML.  
Thus, it is natural to consider that the deviation from the global 
limit is controlled by the variance of $\xi_P$.

\begin{table}[b]
\caption{
The leading $N$ estimate of the ${\cal F}$ in POW$_\alpha$. \label{foliation}}
\begin{tabular}{cccccccccc}
$D$ & $\alpha=0$& $\frac{1}{2}$ & $1$ & $\frac{3}{2}$ & 2  &$\frac{5}{2}$ & 3 & $\frac{7}{2}$ & $\infty$ \\
\tableline
\footnotesize 
1 &$0$&  $\ln N/4N$ & $\pi^2/12\ln^2N$  & $\zeta(3)/2\zeta^2(\frac{3}{2}) $ &
$ \cdots$ &$ \cdots$ &$ \cdots $&$ \cdots$ &$\frac{1}{3}$\\
2 & $0$&  $1/8N$ & $\ln N/4N$ & $\pi^2/96\sqrt{N}$ & $\zeta(3)/2 \ln^2 N$ &$\zeta(4)/8\zeta^2(\frac{3}{2})$ &
$ \cdots$ &$ \cdots $&$\frac{1}{9}$\\ 
3& $0$& $1/24N$&$1/3N$&$\ln N/4N$&$\pi^2/36N^{\frac{2}{3}}$&$\zeta(3)/48N^{\frac{1}{3}}$
& $\pi^4/240\ln^2N$&$ \zeta(5)/24\zeta^2(\frac{3}{2})$ &$\frac{1}{27}$
\end{tabular} 
\label{analyticsuppressionfactors}
\end{table}
The $\xi_P$ is an weighted sum of maps of the form
\begin{eqnarray}
\xi_P=\sum_{Q \in \Lambda} \left( W_{PQ} -\frac{1}{N} \right) f(x_Q) \label{xip}.
\end{eqnarray}
where $W_{PQ}$ is the couplings in (\ref{POW}), (\ref{EXP}), (\ref{cmlkappa}).
If the spatial correlation between the maps are negligible, the variance of $\xi_P$
may be estimated at each time $t$ as
\begin{eqnarray}
 \langle \xi_P^2  \rangle_\Lambda  &\equiv& \langle (h_P-h)^2  \rangle_\Lambda
\approx  {\cal F} \langle (f_P - h )^2  \rangle_\Lambda
 \nonumber \\
{\cal F} &\equiv&  \sum_Q  (W_{PQ})^2 - \frac{1}{N} \label{calF},
\end{eqnarray}
where $\langle \cdots \rangle_\Lambda $ denotes the average over the lattice  
and $\sum_Q W_{PQ}=1$ is used.
The factor ${\cal F}$ represents the suppression of the variance of the $\xi_P$ 
by taking the weighted mean of map values over the lattice $\Lambda$.
At the global limit, $W_{PQ} \rightarrow 1/N$ and ${\cal F}\rightarrow 0$ (strictly no variance).
For intermediate couplings and large $N$ the factor $1/N$ may be neglected
and ${\cal F}$ is solely determined by the couplings. 
Combined with the above consideration let us propose a working hypothesis that 
{\it PM's occur universally in all non-local models when the factor ${\cal F}$ is the same} 
and put it under scrutiny below.

In CML$_\kappa$, the suppression factor ${\cal F}$ is simply ${\cal F}=1/K -1/N$.
This succinctly explains the observation that PM's occur with the same strength at common $K$ 
in all $D$ and uniformly diminish with decreasing (increasing) $K$ (${\cal F}$).

In POW$_\alpha$ the ${\cal F}$  is given by  
\begin{eqnarray}
 {\cal F}_D^{(\alpha)}=\frac{1}{(1+S^{(\alpha)}_D)^2}\left(1+\sum_{\rho=1}^{\rho_{\max}}             
  \frac{n_{\rho,D}}{\rho^{2\alpha}}\right) - \frac{1}{N}. 
\label{POWsuppfac}
\end{eqnarray}
The leading $N$ estimates for ${\cal F}_D^{(\alpha)}$ are tabulated in Table.~\ref{analyticsuppressionfactors}. 
We find in particular  ${\cal F} \approx \log N/4N$ at $\alpha=1/2,1,3/2$ for $D=1,2,3$ respectively.  
This gives a prediction that the PM's would be universal among 
POW$_{\alpha=1/2}^{D=1}$, POW$_{\alpha=1}^{D=2}$ and POW$_{\alpha=3/2}^{D=3}$. 

This is indeed the case; the full strength PM's, the same with those in GCML, are realized in all the three. 
The ${\cal F}_D^{(\rho_0)}$ in EXP$_{\rho_0}$ may be obtained by substituting 
$S_D^{(\rho_0)}$ and $w^2_{\rho,\rho_0}$ to $S_D^{(\alpha)}$ and $1/\rho^{2\alpha}$ respectively. 

\begin{figure}[t]
\begin{center}
\leavevmode
\epsfxsize=110mm
\epsfysize=85.74mm
\epsfbox{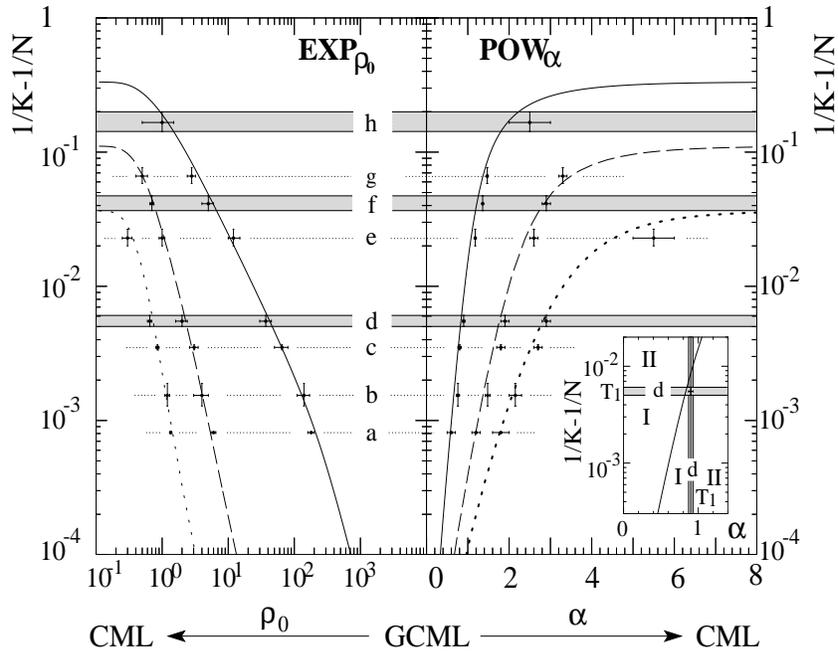}
\vspace{5mm}
{\footnotesize\caption{
The comparison between POW$_\alpha$ and CML$_\kappa$ (right)
and EXP$_{\rho_0}$ and CML$_\kappa$ (left) 
with respect to eight marked changes (a-h) of the PM's.
For $D=1,2,3$ respectively the ${\cal F}_D$ is shown by 
the solid, dashed, and dotted curves and $N=51^2, 51^2, 13^3$.
The inset illustrates the method for the case of d (T$_1$) in $D=1$.
The $\alpha$ ($\rho_0$) axis is in the normal (logarithmic) scale.
\label{modelcomparison}}}
\end{center}
\end{figure}

In Fig.~\ref{modelcomparison} we compare POW$_\alpha$ (EXP$_{\rho_0}$) with CML$_\kappa$
 in the right (left). The inset illustrates the case of the transition point T$_1$(d) in $D=1$
as an example.  The curve is ${\cal F}$ for POW$_\alpha$ in (\ref{POWsuppfac}).
The measured $\alpha$ at T$_1$ gives the vertical band taking account for the ambiguity 
in judging the MSD curve pattern.  
Hence,  the crossing of the curve and the vertical band gives the estimate of ${\cal F}$ in POW$_\alpha$ 
at its T$_1$ in $D=1$. 
On the other hand, ${\cal F}$ is universal over $D$ in CML$_\kappa$.
The  ${\cal F}$ at T$_1$ of CML$_\kappa$ gives the horizontal band, again 
counting for the ambiguity and averaged over $D$.
Thus, the vertical axis is used for both ${\cal F}$'s, that in POW$_\alpha$ and that in CML$_\kappa$.
If both models share exactly the same ${\cal F}$ at T$_1$, the curve will pass through the crossing junction
of the horizontal and vertical bands. In this example, the curve crosses 
the vertical band slightly above the junction
and the estimated ${\cal F}$ are $(8.5 \pm 1.5) \times 10^{-3}$ 
and $(5.5 \pm 0.5) \times 10^{-3}$ 
in POW$_\alpha$ and CML$_\kappa$ 
respectively.  Or, one can predict the $\alpha$ at T$_1$ in 
POW$_\alpha$ from $K$ at T$_1$ in CML$_\kappa$ using
the ${\cal F}$ curve of POW$_\alpha$. 
The prediction is $0.83 \pm 0.02$ to be compared with the measured $0.90 \pm 0.03$. 
In the overall comparison, only the junctions are shown by error-bars. 
We find that the hypothesis remarkably works with respect to all the marking points and in $D=1,2,3$
for ${\cal F}$ ranging from $10^{-4}$ to $O(1)$.

A few remarks are in order. 

(i) The curves in POW$_\alpha$ agree approximately with each other 
after a scale transformation $1:1/2:1/3$ in $\alpha$ up to ${\cal F}_D \approx 2 \times 10^{-2}$.
This extends the rule obtained by the leading $N$ calculation. 
The same strength PM's occur up to the marking point e if $\alpha$ is 
in the ratio of the system dimensionality, just like the universal PM's at the same $K$ in CML$_\kappa$.  

(ii) The horizontal bands exhibit three transition points observed in 
CML$_\kappa$ using refined neighborhood. 
In other models with the coarse neighborhood, the T$_3$ is missing in $D=2$ and 
both T$_2$ and T$_3$ are missing in $D=3$. (See Fig.~\ref{barchartpow}
for POW$_\alpha$). 
The curves of ${\cal F}$ explain the difference succinctly; 
they are constrained by the limiting values $1/3^D$ so they can pass through 
only the lowest two (one) bands in $D=2(3)$. We have numerically checked 
that the missing transition points are retrieved in both POW$_\alpha$ and EXP$_{\rho_0}$
with the refined neighbors.

(iii) Our estimate of ${\cal F}$ is based on an approximation that the spatial correlation is negligible.
This is a legitimate approximation. Firstly, we note that the even at the formation of 
cluster attractors such as $p3c3$ MSCA and $p3c2$ states, the spatial distribution of maps 
does not show any visible clustering. To avoid a confusion we stress that the clustering of maps 
is in the map values and not in the spatial distribution.  We have checked that over the whole turbulent 
regime of the three models no visible spatial clusters are formed. 
Furthermore we have numerically checked that the estimate (\ref{calF}) works 
remarkably well in three models. 
For $D=1-3$ and over $\varepsilon=0-0.1$ the measured ratio $\langle\xi^{2}_{P}\rangle_{\Lambda}/\langle(\delta f_{P})^{2}\rangle_{\Lambda}$ plotted versus ${\cal F}$ both in double logarithmic scale lies 
on a straight line for ${\cal F}$ from $10^{-5}$ to $O(1)$, which fully covers the range of ${\cal F}$ in Fig.~\ref{modelcomparison}. There turns out some spread of data points only in the $\varepsilon$ region for the $p3c2$ cluster attractor. This is due to the variation of the ratio of map populations in two clusters when formed from different initial configurations.\\ \\

\noindent
{\bf 4. Conclusion}\\

In this note we have focused our attention to the recently found periodicity manifestations in the 
turbulent regime of GCML. We have conducted an extensive statistical analysis 
in three non-locally coupled map lattices over $D=1,2,3$  
and examined to what extent they depend on the non-locality of the models.   
We have noted that the essential deviation of the non-local CML
from the GCML stems in the variance of the local mean field around the overall mean field.
We have analytically estimated the suppression factor ${\cal F}$ of the variance 
under an approximation that the spatial correlation of maps in the turbulence 
regime is negligible and checked that this ${\cal F}$ remarkably 
agrees with the numerical result. 
We have found a salient universally that, irrespective of the difference in construction and
the dimension of the lattice, the periodicity manifestations occur at the same strength to a good approximation 
once ${\cal F}$ is the same.\\ \\

\noindent
{\bf Acknowledgements}\\

One of authors (T.S.) especially thanks Wolfgang Ochs for encouragement and reading the manuscript, 
and Max-Planck Institut f\"ur Physik, M\"unchen for the warm hospitality during his visit. 
Our thanks also go to Mario Cosenza for communication and Kengo Kikuchi for collaborating with us at the early stage of this work. 


\end{document}